\documentclass[graybox]{svmult}

\usepackage{mathptmx}       
\usepackage{helvet}         
\usepackage{courier}        
\usepackage{type1cm}        
\usepackage{makeidx}         
\usepackage{graphicx}        
\usepackage{multicol}        
\usepackage[bottom]{footmisc}
\usepackage{epsfig}

\makeindex             

\begin{document}

\title*{Nuclear Track Detectors. Searches for Exotic Particles.}
\author{Giorgio Giacomelli$^1$ and Vincent Togo$^2$}
\institute{$^1$University of Bologna and INFN Bologna, Italy; \email{giacomelli@bo.infn.it} \\
           $^2$INFN, Sezione di Bologna, Italy; \email{togo@bo.infn.it}}

\maketitle

\abstract{We used Nuclear Track Detectors (NTD) CR39 and Makrofol
for many purposes: i) Exposures at the SPS and at lower energy accelerator heavy ion beams for calibration purposes 
and for fragmentation studies. ii) Searches for 
GUT and Intermediate Mass Magnetic Monopoles (IMM),  
nuclearites, Q-balls and strangelets in the cosmic
radiation. The MACRO experiment in the Gran Sasso underground lab, with $\sim$1000 m$^2$ of CR39 detectors
(plus scintillators and streamer tubes),
established an upper limit for superheavy GUT poles at the level
of $1.4 \times 10^{-16}$ cm$^{-2}$ s$^{-1}$ sr$^{-1}$ for $4 \times 10^{-5}$$<$$\beta$$<$1.
The SLIM experiment at the high altitude Chacaltaya lab (5230 m
a.s.l.), using 427 m$^2$ of CR39 detectors exposed for 4.22 y, gave an
upper limit for IMMs of $\sim$$1.3 \times 10^{-15}$ cm$^{-2}$ s$^{-1}$ sr$^{-1}$. 
The experiments yielded interesting upper limits also on
the fluxes of the other mentioned exotic particles. 
iii) Environmental studies, radiation monitoring, neutron dosimetry. }


\section{Introduction}\label{sec:1}
Nuclear Track Detectors are used in many branches of science and technology \cite{Durrani}. The isotropic 
poly-allyl-diglycol carbonate polymer, commercially known as CR39, is the most sensitive NTD; also Makrofol  
and Lexan polycarbonates are largely employed.\par
A nuclear track detector records the passage of highly ionizing particles via their Restricted Energy  
Loss  (REL).  The  latent  damage trail formed in NTDs may be enlarged by a suitable chemical etching and made visible under an 
optical microscope. The latent track develops into a conical-shaped etch-pit (Fig. 1) when the etching velocity along the particle 
trajectory ($v_T$) is larger than the bulk etching velocity of the material ($v_B$) (\cite{Nikezic}).
The sensitivity of NTDs crossed by particles with constant energy loss is characterized by the ratio $p = v_T/v_B$ (reduced etch 
rate) which is determined by measuring the bulk etch rate $v_B$ and either the etch-pit diameter or height. 
Two methods were used to determine $v_B$. The first is the common one based on the mean thickness difference  
before and after etching. The second method is based on both cone height and base diameter 
measurements of the etched tracks.
\par The measured track diameter D and track length L$_e$ are expressed in terms of the velocities $v_T$ and $v_B$\\
\begin{equation}
D = 2v_{B}t\sqrt{\frac{(v_{T}-v_{B})}{(v_{T}+v_{B})}}
\label{eq:2}
\end{equation}

\begin{figure}[t]
\begin{center}
\includegraphics[height=3.3cm]{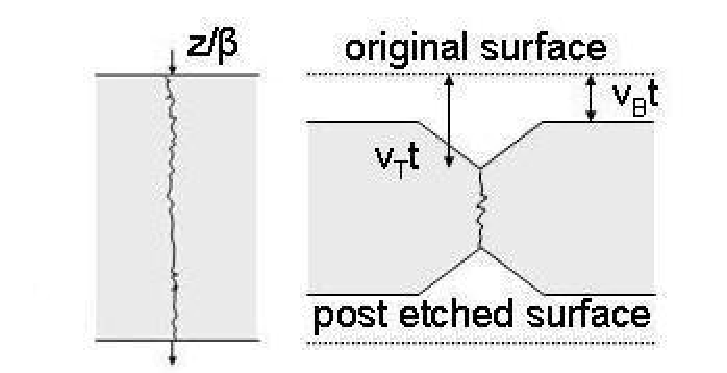}
\hspace{-0.5cm}
\includegraphics[height=3.4cm]{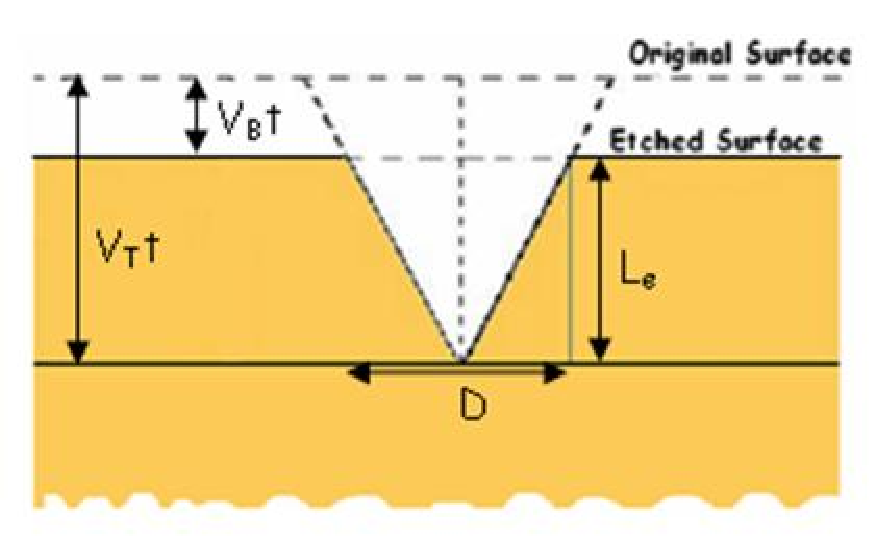}
\caption{Sketch of: (a) the latent track formation by a charged particle passing in a NTD. (b) Situation after 
etching. (c) Parameters of an etched track for a normally incident fast ion  in a NTD. }
\label{fig:1}       
\end{center}
\vspace{-0.8cm}
\end{figure}

\begin{equation}
L_{e} = (v_{T}-v_{B}) t
\end{equation}
from which one obtains
\begin{equation}
p = \frac{v_{T}}{v_{B}} = 1 + \frac{L_{e}}{v_{B}t} = \frac{1+(\frac{D}{2v_{B}t})^{2}}{1-(\frac{D}{2v_{B}t_{}})^{2}}
\label{eq:6}
\end{equation}
Experiments in different fields require an accurate detector calibration \cite{Uchihori, Kodaira}. 
More than 4000 m$^2$ of CR39 
detectors were used in the MACRO and SLIM experiments which searched for new massive particles 
in the cosmic radiation (magnetic monopoles, nuclearites, Q-balls) [5-13].
\par
In this note will be summarized the technical work on NTDs and results on fragmentation studies, on the 
search for MMs, Nuclearites and Q-balls in the cosmic radiation, underground or at high altitudes, and environmental 
monitoring.

\begin{figure}
\begin{center}
\includegraphics[height=3.2cm]{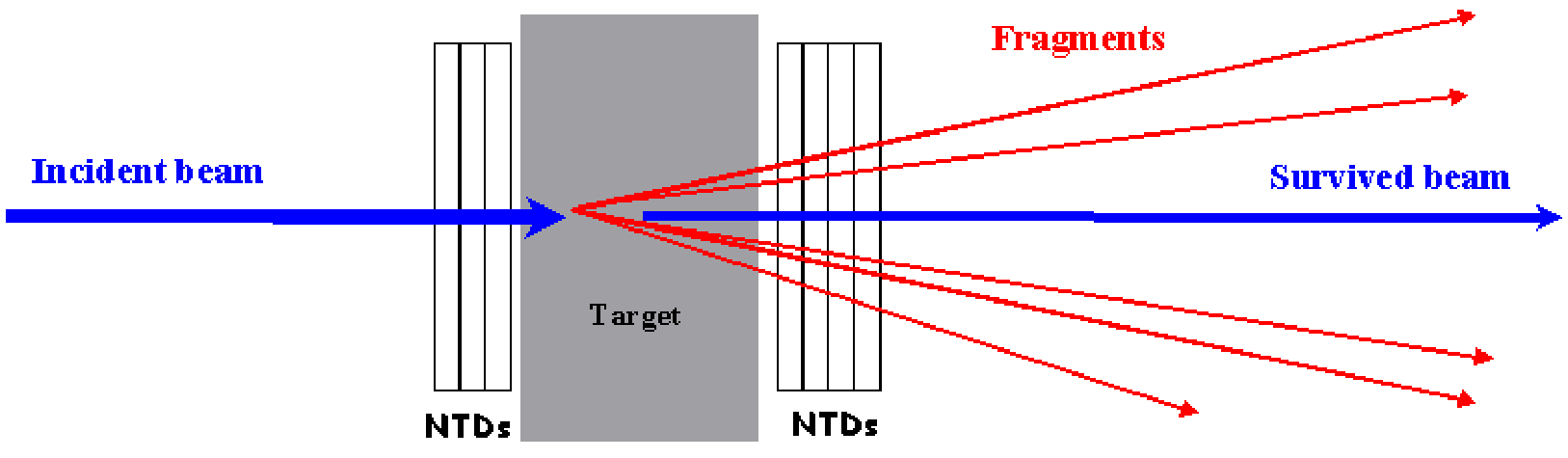}
\vspace{-0.2cm}
\caption{Exposure set-up for the calibration of CR39 and Makrofol NTDs.}
\label{fig:2}       
\end{center}
\vspace{-0.5cm}
\end{figure}

\vspace{-0.5cm}
\section{Experimental. Calibrations}\label{sec:2}

Stacks composed of CR39 and Makrofol foils of size 11.5$\times$11.5 cm$^2$ with several targets were exposed to 
158 A GeV In$^{49+}$ ions in 2003 at the CERN-SPS, at normal incidence and a total ion density of $\sim$2000 /cm$^2$. The detector 
foils downstream 
of the target recorded the beam ions as well as their nuclear fragments [14-16] 
see Fig. 2. The CR39 polymer sheets were manufactured by Intercast Europe Co., Parma, Italy, where a scientific 
production line was set up in order to achieve a lower detection threshold, a higher sensitivity in a larger range of energy losses, 
a high quality of the post-etched surfaces after prolonged etching [17,18]. 
The Makrofol detectors were manufactured by Bayer A.G., Germany.   
\par After exposures, CR39 and Makrofol foils located after the target were etched 
in 6N NaOH + 1\% ethyl alcohol at 70$^\circ$C for 40 h and 6N KOH + 20\% ethyl alcohol at 50$^\circ$C for 8 h that are the
optimum etching condition for CR39 and Makrofol, respectively.
 The addition of ethyl alcohol in the etchant 
improves the etched surface quality, reduces the number of surface defects and background tracks, increases the bulk 
etching velocity, speeds up the reaction, but raises the detection threshold 
\cite{Manzoor2007,Balestra2007}.
 The etching was performed in a stainless steel tank equipped with internal thermo-resistances and a motorized stirring head.
A continuous stirring 
was applied; the temperature was stable to within $\pm$ 0.1$^{\circ}$C.\par

\begin{figure} 
\begin{center}
\includegraphics[height=4cm]{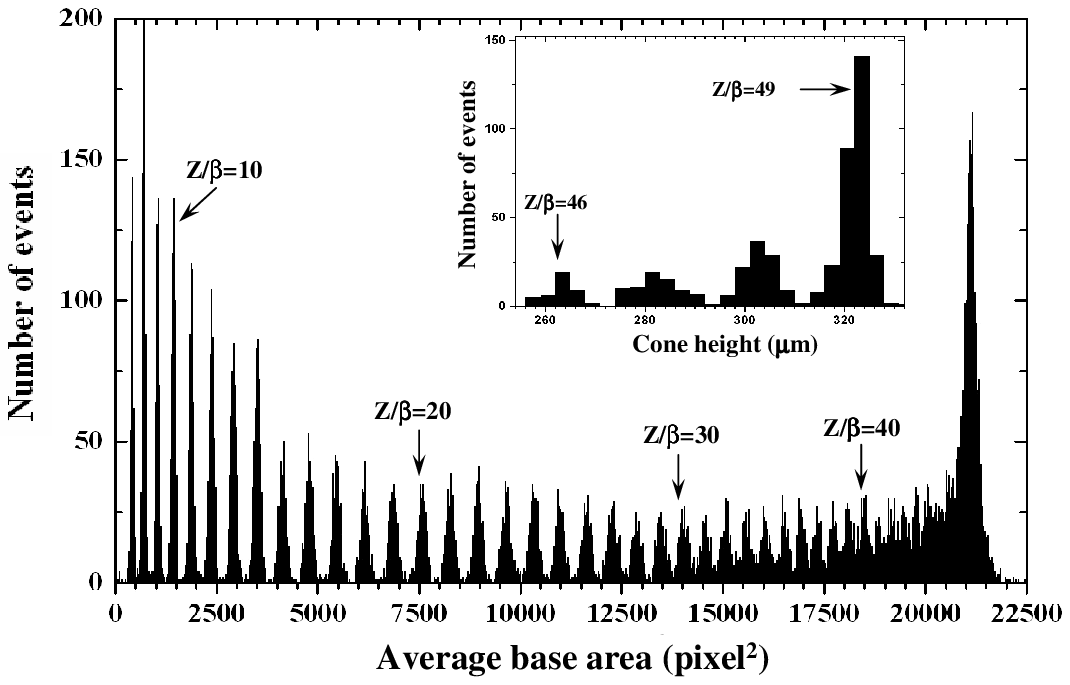}
\includegraphics[height=4cm]{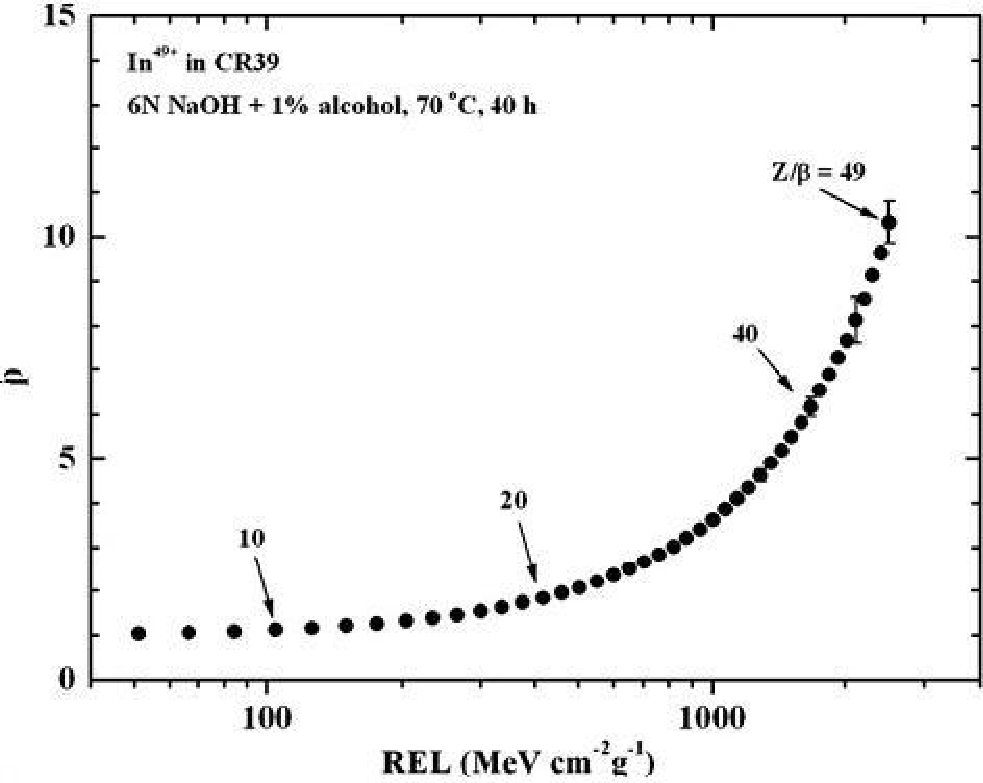}
\caption{(a) Base area distributions of etched cones in CR39 from 158 A GeV In$^{49+}$ ions and their fragments.
In the insert are shown the cone height distribution for $46\le Z/\beta\le 49$.  
(b) p vs REL for CR39; statistical standard
deviations are shown at  $Z/\beta = 40, 45, 49$; for $Z/\beta< 37$ the errors are smaller than the size of the points.  }
\label{fig:3}       
\end{center}
\vspace{-0.5cm}
\end{figure}

     Fig. 3a shows the etch-pit base area distribution for indium ions and their fragments in CR39 measured with the Elbek 
image analyzer system \cite{Noll}; averages were computed from measurements made on the ''front sides'' of two detector 
sheets. The fragment peaks are well separated, from $Z/\beta\sim$7  to  45; the charge resolution for the average of two 
measurements is 
$\sigma_Z\sim$0.13e at $Z/\beta\sim$15. The resolution close to the indium peak (Z = 49) 
can be improved by measuring the heights of the etch-pit cones 
\cite{Giacomelli98}.

They were measured with a Leica microscope coupled to a CCD camera and a video monitor; the L$_e$  
distribution is shown in the insert in  Fig. 3a \cite{Balestra2007}.
For each nuclear fragment we computed the REL and the reduced etch rate 
p using  Eq. 3; p vs REL is plotted in Fig. 3b; the CR39 detection threshold is at REL$\sim$50 MeV cm$^2$ g$^{-1}$ 
(corresponding to a relativistic fragment with $Z/\beta$$\sim$7).
\par Measurements of {\it Makrofol detectors} exposed to Pb ions and their fragments yield fragmentation peaks well separated from 
$Z/\beta\sim$51 to $\sim$77. The threshold is at $Z/\beta$$\sim$50; the charge resolution for 2 face measurement is 
$\sigma_Z\sim$0.18e at $Z/\beta\sim$55.

\begin{figure}[t]
\begin{center}
\includegraphics[height=5.5cm]{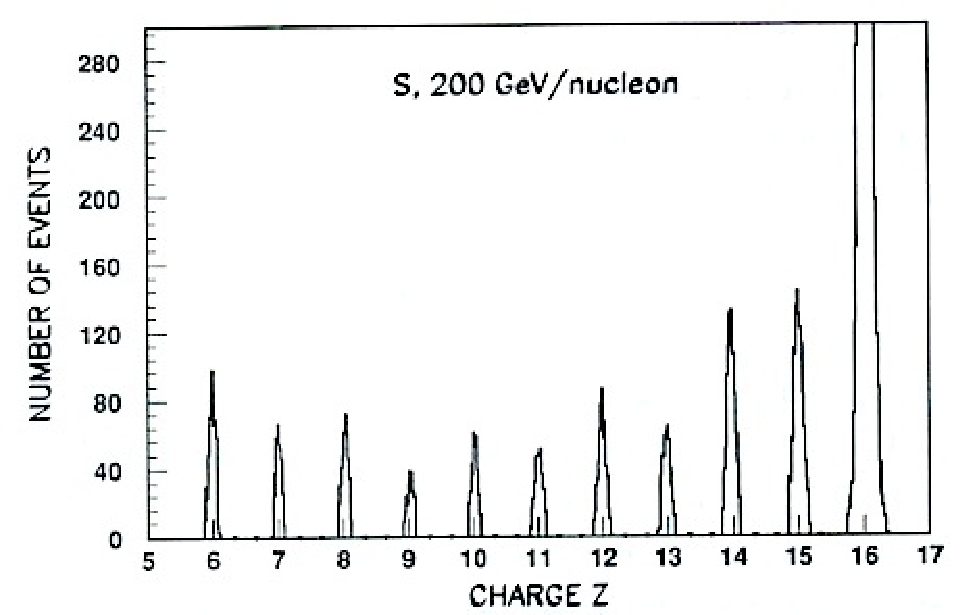}
\caption{Search for nuclear fragments with fractional charges in CR39 detectors (12 measurements were made on the same track)}
\label{fig:4}       
\end{center}
\vspace{-0.5cm}
\end{figure}

\par
Fig. 4 shows the results of repeated precision measurements on the same track (12 times) in CR39 detectors exposed to 200 
GeV/nucleon S ions and their fragments. The charge resolution is adequate to allow a search for fragments with fractional charges. 
The limits on fractional charge fragments are at the level of $10^{-4}$ \cite{Cecchini93}.

\vspace{-0.5cm}

\section{Fragmentation cross sections}\label{sec:3} 
The availability of ion beams at CERN, BNL and Chiba (HIMAC) made possible to investigate the projectile  
fragmentation on different targets and for rather different projectile energies. 
The total charge changing cross sections were determined from the survival
fraction of ions using the following relation
 \begin{equation}
 \sigma_{tot} = \frac {A_T \ln (N_{in} / N_{out})}{\rho~ t ~N_{Av}}
\end{equation}
where $A_T$ is the nuclear mass of the target; $N_{in}$ and $N_{out}$ are the
numbers of incident ions before and after the target, respectively; $\rho$
(g/cm$^3$) is the target density; $t$ (cm) is the target thickness  and 
$N_{Av}$ is Avogadro number.\par
At low energies the total fragmentation cross sections are essentially energy independent and are in agreement with 
semi-empirical 
formula \cite{Bradt}. At high energies the fragmentation cross section depends on 
the target mass as 
$\sim$A$_T^{1/3}$. The partial fragmentation cross sections increase with decreasing $\Delta Z$ and the cross sections leading to 
even $Z$ fragments are slightly larger than those leading to odd $Z$ \cite{Cecchini2008,Giorgini}.

\vspace{-0.5cm}
\section{Searches for Magnetic Monopoles}\label{sec:4}  

GUT theories of the electroweak and strong
interations predict the existence of superheavy MMs
produced in the Early Universe when the
GUT gauge group breaks into separate groups, one of which is
U(1): 
\begin{equation}
\footnotesize
    \begin{array}{ccccc}
        {} & 10^{15}\ GeV & {} & 10^{2}\ GeV & {} \\
        SU(5) & \longrightarrow & SU(3)_{C}\times \left[
        SU(2)_{L}\times U(1)_{Y}\right] & \longrightarrow &
        SU(3)_{C}\times U(1)_{EM} \\
       {} & \small10^{-35}s & {} & \small10^{-9}s & {}
    \end{array}
\end{equation}
MMs should be generated as topological point defects in the GUT phase
transition $SU(5)\longrightarrow U(1)_Y$, about one pole for each
causal domain. In standard cosmology this leads to too many poles
(\emph{the monopole  problem}). A rapid expansion of the early Universe
(\emph{inflation}) would defer the
GUT phase transition; in the simplest inflation version the number
of generated MMs is small. However if there was a reheating
phase one may have MMs produced in
high energy collisions, like
$e^{+}e^{-}\rightarrow M\bar{M}$. 
\par Fig. 5 shows a sketch of the energy loss of a MM in liquid H.
\par The structure of a GUT MM: a very small
core, an electroweak region,  a confinement region, a
fermion--antifermion condensate (which may contain 4--fermion
baryon--number--violating terms); for $r\geq$ few $fm$ a GUT pole
behaves as a point particle generating a field
$B=g/r^{2}$, Fig. 6 \cite{Derkaoui}.

\begin{figure}[t]
\begin{center}
\includegraphics[height=5.2cm]{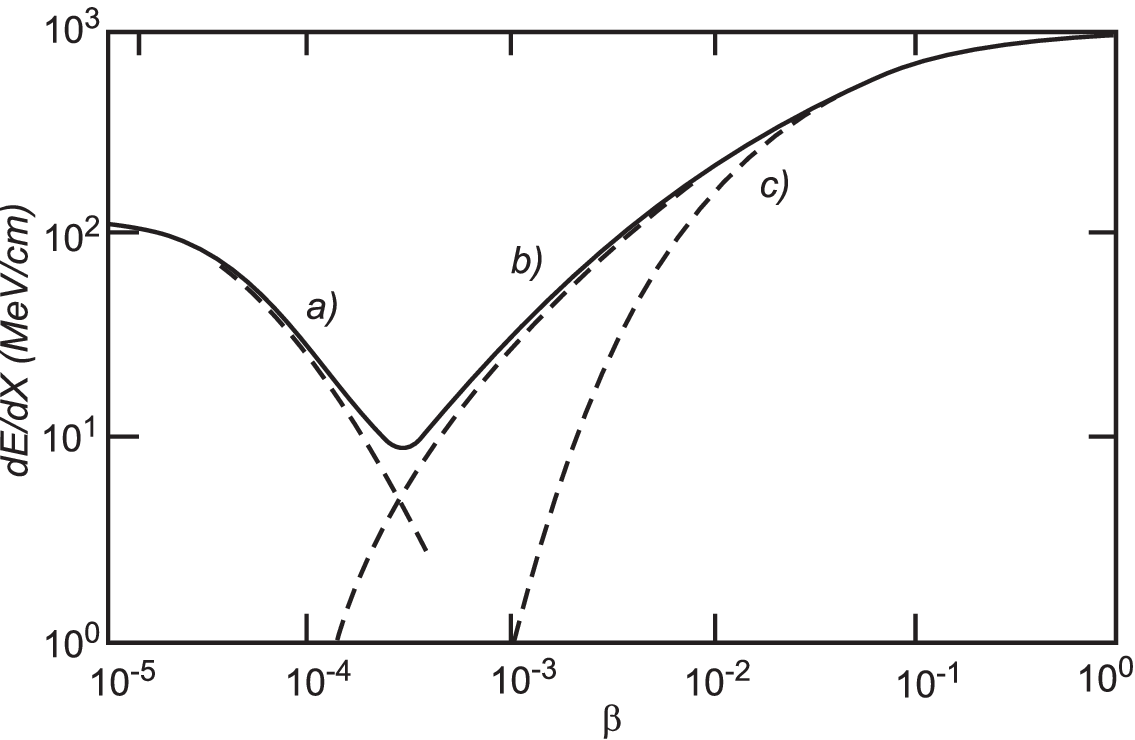}
\caption{The energy losses of $g=g_D$ MMs in liquid hydrogen due to (a) elastic collisions; (b) excitation and Drell 
effect; (c) ionization.}
\label{fig:5}       
\end{center}
\vspace{-0.5cm}
\end{figure}

\begin{figure}[t]
\begin{center}
\includegraphics[height=5cm]{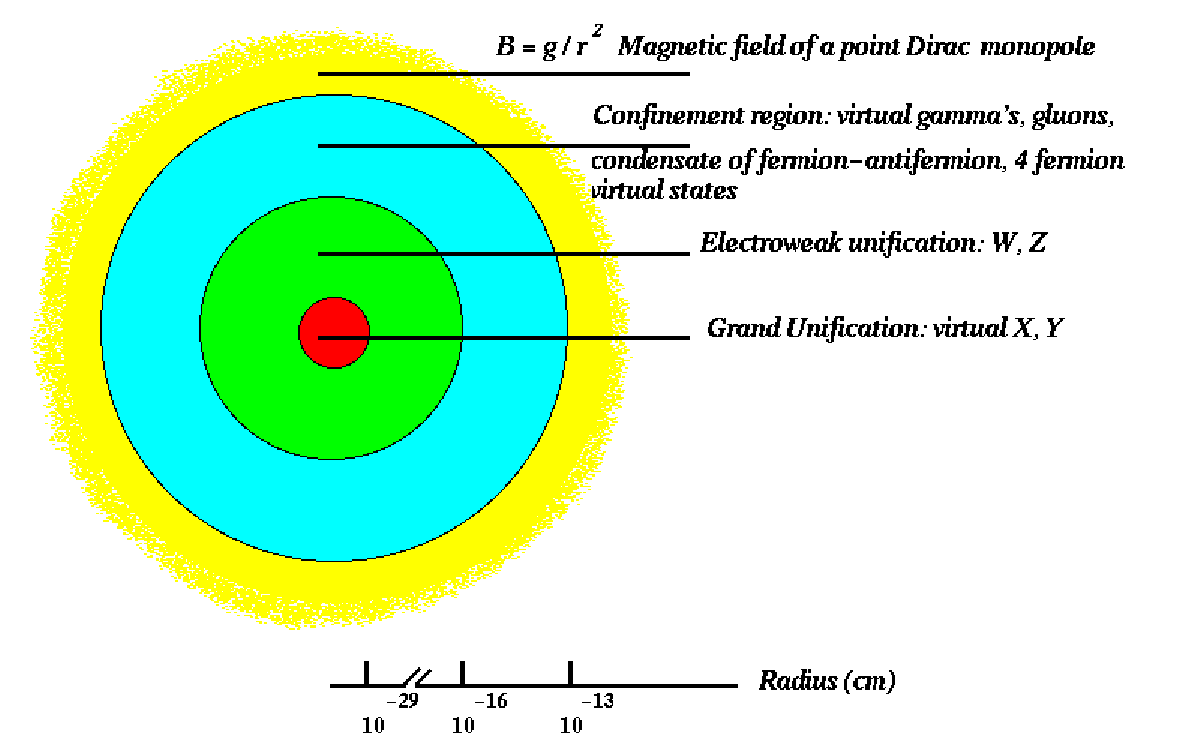}
\caption{Extended picture of a GUT Magnetic Monopole.}
\label{fig:6}       
\end{center}
\vspace{-0.5cm}
\end{figure}

\par A flux of cosmic GUT MMs may reach the Earth with a velocity spectrum
in the range $4 \times 10^{-5}$$<$$\beta$$<$0.1, with possible peaks
corresponding to the escape velocities from the Earth, the Sun and the
Galaxy. Searches in the Cosmic Rays (CR) performed with
superconducting induction devices yielded a 90\%~CL limit of
$2\times$$10^{-14}~$cm$^{-2}$~s$^{-1}$~sr$^{-1}$ independent of $\beta$ \cite{gg+lp}.
\par Direct searches were performed above ground and
underground \cite{Derkaoui, ruzicka, biblio}. MACRO
 made a search with liquid scintillators, limited streamer tubes and
NTDs; the  90\% CL flux limits obtained with the NTDs are shown  in
Fig. 7a. Fig. 7b shows the limits for $g=g_D$ obtained with all the subdetectors; they are at the level of
$1.4\times 10^{-16}$~cm$^{-2}$~s$^{-1}$~sr$^{-1}$ for $\beta > 4
\times 10^{-5}$~\cite{Ambrosio2002a, mm_macro}. The figure shows also the limits from
 the Ohya, Baksan, Baikal, and AMANDA experiments \cite{Experiments}.\par

\begin{figure}[t]
\begin{center}
\includegraphics[height=4.4cm]{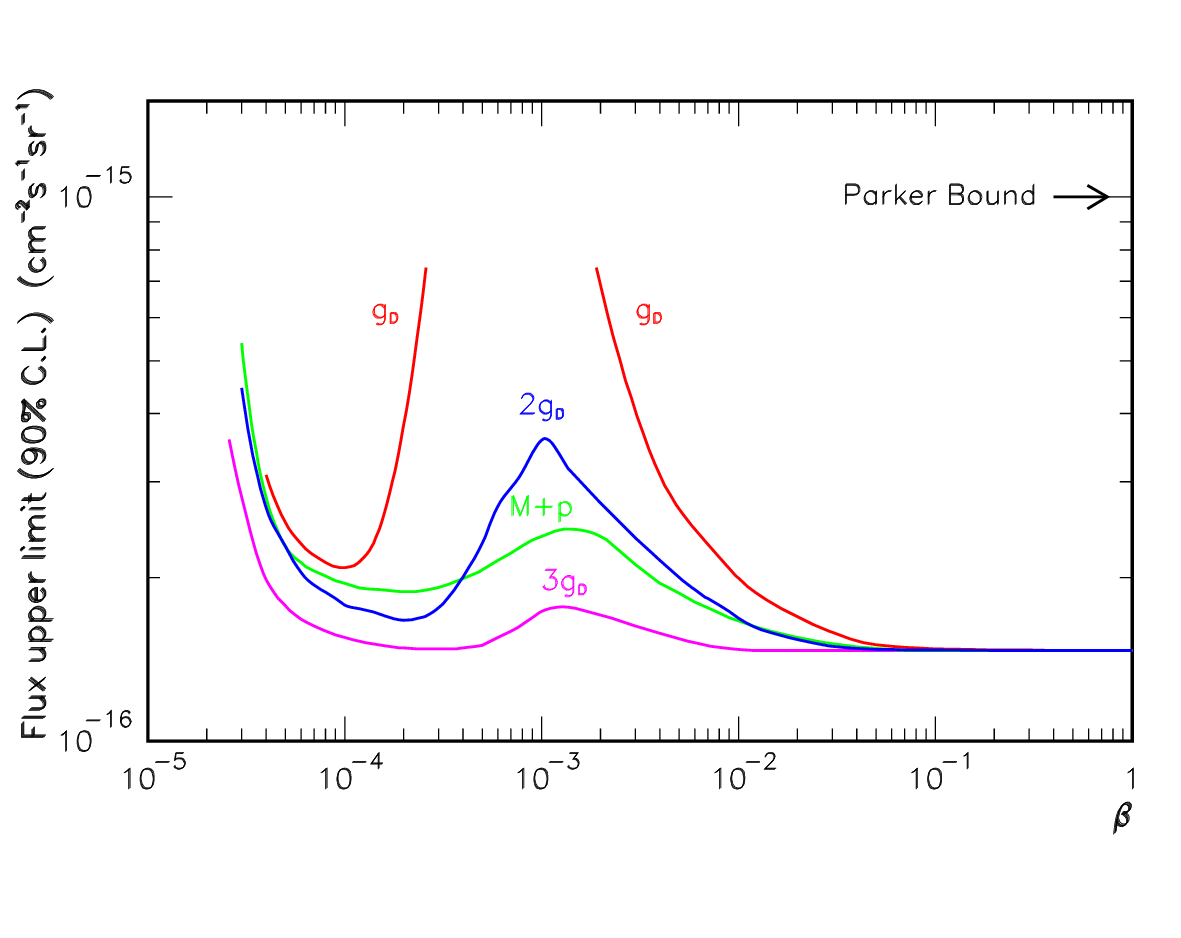}
\includegraphics[height=4.1cm,width=6cm]{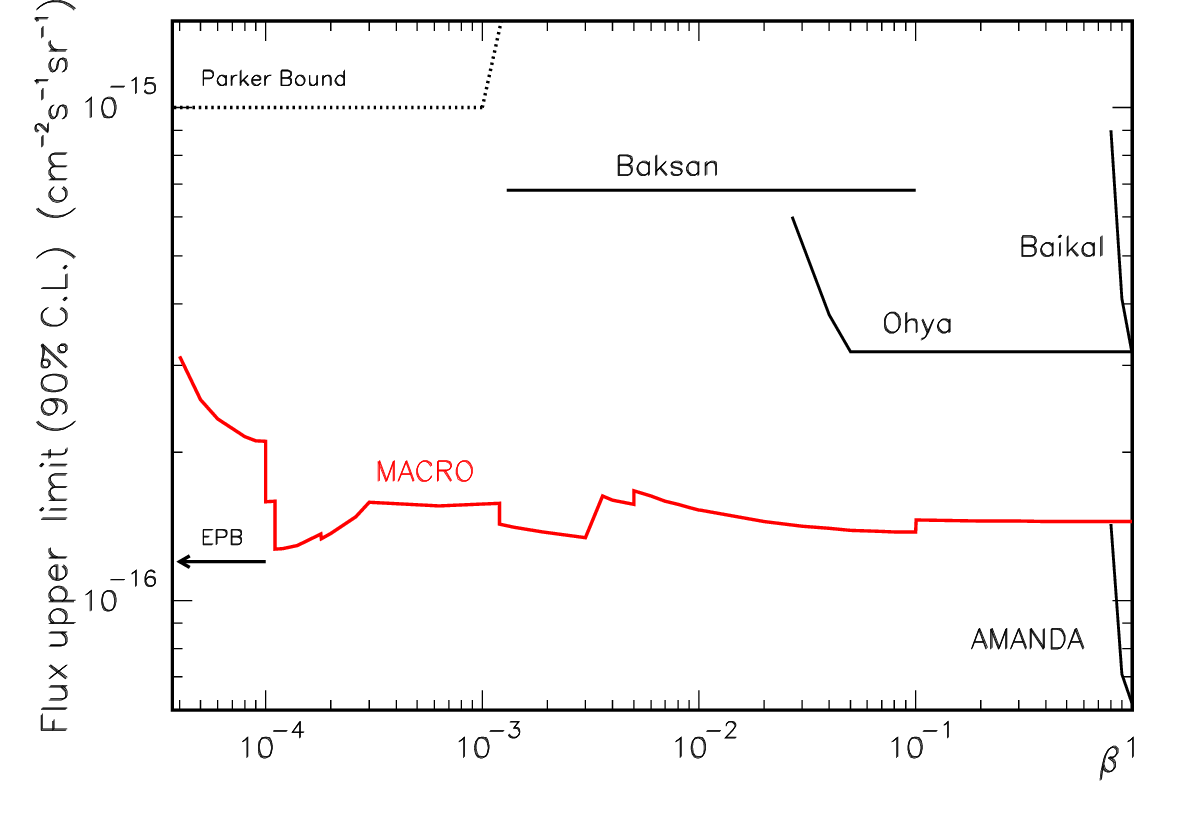}
\vspace{-0.2cm}
\caption{(a) Upper limits (90\%) for an isotropic flux of MMs obtained with the CR39 subdetector of MACRO
for poles with magnetic charge g=g$_D$, 2g$_D$, 3g$_D$ and for M+p composites.
(b) Global limit obtained by MACRO for GUT poles with  g=g$_D$, using all its subdetectors.}
\label{fig:7}       
\end{center}
\vspace{-1cm}
\end{figure}

\begin{figure}
\begin{center}
\includegraphics[height=6cm]{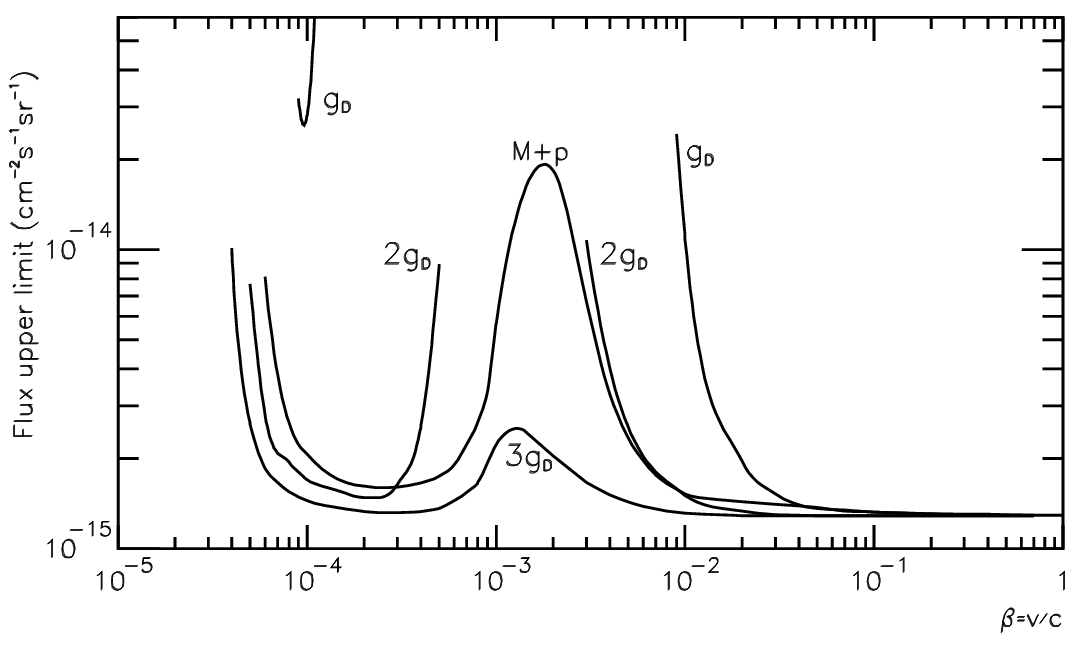}
\vspace{-0.2cm}
\caption{90\% CL upper limits from the SLIM experiment on IMMs in the cosmic radiation.}
\label{fig:8}       
\end{center}
\vspace{-0.5cm}
\end{figure}

\begin{figure}
\begin{center}
\includegraphics[height=2.7cm]{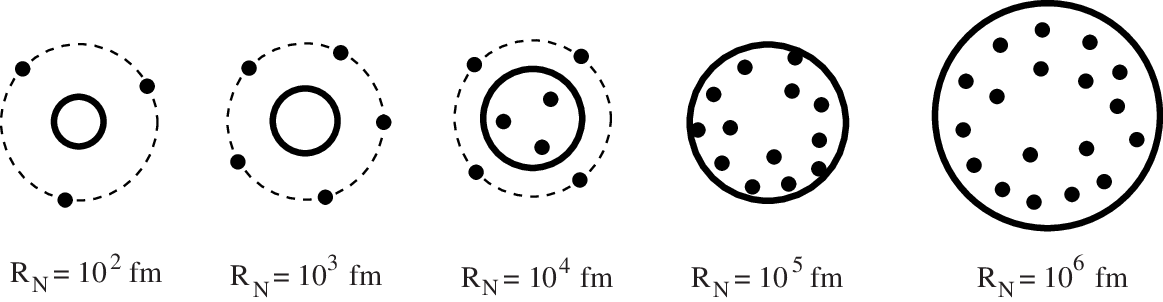}
\caption{Sketch of nuclearite structure: the quark bag radius R$_N$ and the core plus electron system (indicated by dashed
lines); the electrons are indicated by black points.}
\label{fig:9}       
\end{center}
\end{figure}

\par
Indirect GUT MM searches used ancient mica which is a NTD with a
very high threshold. It is assumed that a pole passing through the Earth
captures an Al nucleus and drags it through subterranean mica causing
lattice defects, which survive as long as the mica is not 
reheated. Only small sheets were analyzed ($13.5$ and $18$ cm$^2$),
but they recorded tracks for $4\div9\times 10^8$
y. The flux limits are at the level of $\sim$10$^{-17} ~\mbox{cm}^{-2}~
\mbox{s}^{-1} $sr$^{-1}$ for $10^{-4}<\beta<10^{-3}$~\cite{price}. But these indirect experiments might not be
really so sensitive.\par

{\it Intermediate Mass Magnetic Monopoles} may appear as topological point defects at a later time in the     
Early Universe, f.e. if the GUT group yields the U(1) group of the 
Standard Model in the following two steps:
\begin{equation}
\footnotesize
    \begin{array}{ccccc}
        {} & 10^{15}\ GeV  & {}& 10^{9}\ GeV & \\
        SO(10) & \longrightarrow & SU(4)\times SU(2)\times SU(2) &
        \longrightarrow & SU(3)\times SU(2)\times U(1) \\
        {} & \small10^{-35}s & {} & \small10^{-23}s & {}
    \end{array}   
\end{equation}
\noindent This would lead to MMs with masses of $\sim$$10^{10}$ GeV;   
they would survive inflation, be stable, ``doubly charged'' ($g=2g_D$) 
and do not catalyze nucleon decay~\cite{lazaride}.
The structure of an IMM is similar to that of a
GUT MM, but the core is larger. 
\par
Relativistic IMMs with masses, $10^7$$<$$m_M$$<$$10^{13}$ GeV, may be present in the 
cosmic radiation, and may  be  accelerated to high $\gamma$ factors in one      
domain of the galactic magnetic field. Detectors at the Earth surface may detect downgoing IMMs if 
$m_M$$>10^5$ GeV \cite{Derkaoui}; 
lower mass MMs may be detected at high mountain altitudes,
in balloons and in satellites. 
\par
SLIM at 5230 m a.s.l. \cite{Balestra2008} was based on
427 $m^2$ of      
CR39 and Makrofol detectors exposed for 4.27 years to the CR. The
detectors were organized in modules of 24$\times$24 $cm^2$, each
consisting of 3 layers of CR39 
interleaved with 3 layers of Makrofol and 1 mm Al absorber. Each module was packed in an
aluminized polyethylene envelope at 1 atm of dry air to prevent the
CR39 loss in sensitivity at a reduced oxygen content in the air (0.5 atm). 
The 90\% CL flux upper limits for downgoing IMMs with
$g=g_D,~2g_D,~3g_D$ and M+p are plotted in Fig. 8 
vs $\beta$ ($2g_D$ is the theoretically preferred value).

\begin{figure}[t]   
\begin{center}
\includegraphics[height=7cm]{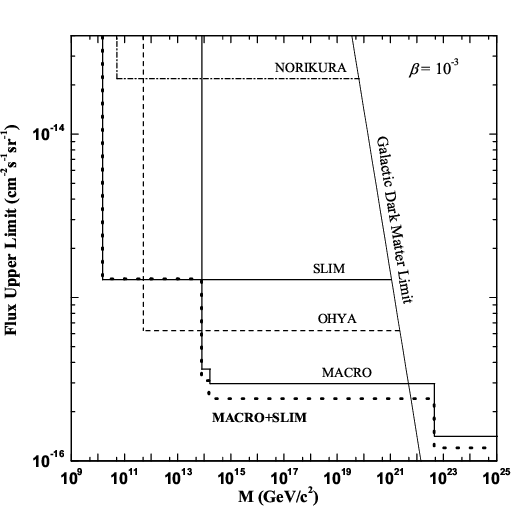}
\caption{Flux limits at 90\% CL for downgoing nuclearites versus mass.}
\label{fig:10}       
\end{center}
\vspace{-0.5cm}
\end{figure}

\begin{figure}
\begin{center}
\includegraphics[height=3.5cm]{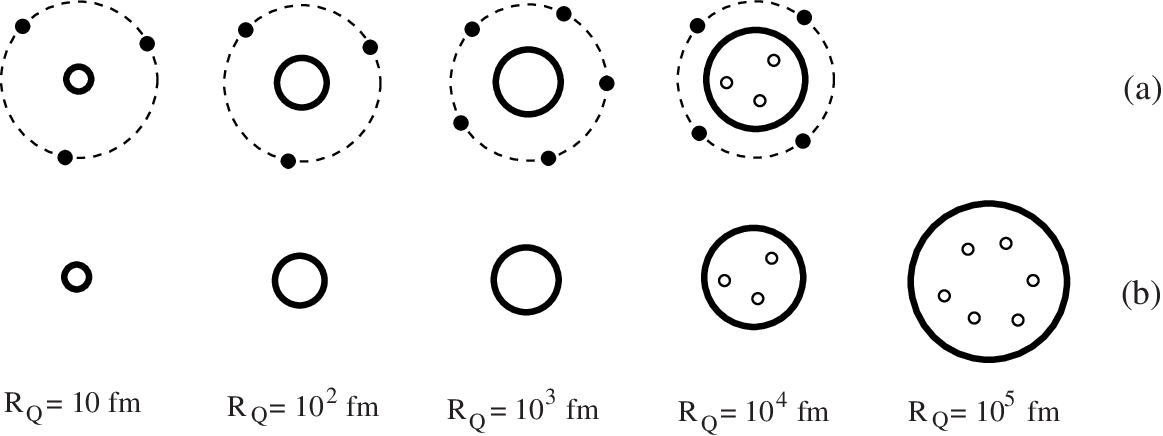}
\caption{Sketch of Supersymmetric Q-ball structure: (a) for SECS and (b) for SENS. The black points are electrons, the empty dots 
are s-electrons.}
\label{fig:11}       
\end{center}
\vspace{-0.5cm}
\end{figure}

\begin{figure}   
\begin{center}
\includegraphics[height=6.5cm]{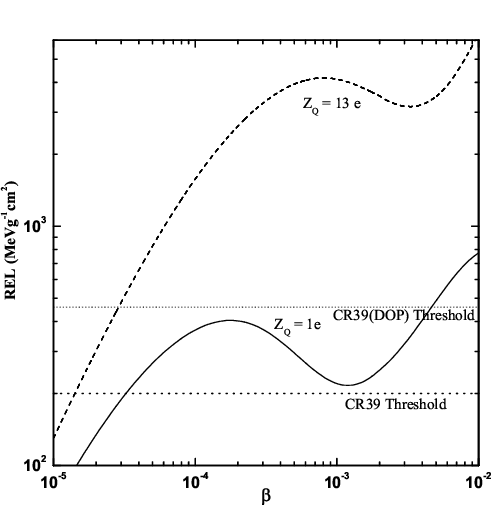}
\caption{REL vs $\beta$ for downgoing charged Q-balls (SECS).}
\label{fig:12}       
\end{center}
\vspace{-0.5cm}
\end{figure}

\vspace{-0.5cm}
\section{Searches for Nuclearites, Strangelets, Q-balls}\label{sec:5}

{\it Strange Quark Matter} (SQM) consists of aggregates of \textit{u,
  d} and \textit{s} quarks in almost equal proportions (the number of
\textit{s} quarks is lower than the number of \textit{u} or
\textit{d} quarks and SQM should have a relatively small positive integer
charge). The overall neutrality of SQM is ensured by an electron cloud
which surrounds it, forming a sort of atom, see
Fig. 9 \cite{Derkaoui, nucleariti, Cecchini2008b, macro-nucl}.  SQM has a constant density
$\rho_N = M_N /V_N\simeq 3.5 \times 10^{14}$~g~cm$^{-3}$, slightly
larger than that of atomic
nuclei, and it should be stable for all baryon numbers between ordinary heavy nuclei and neutron stars. 
SQM with baryon number $A<10^6-10^7$ are called
``\emph{strangelets}''; the word ``\emph{nuclearite}'' was  
introduced to indicate larger lumps of SQM, which may be present in the
Cosmic Radiation \cite{Medinaceli, nucleariti, Cecchini2008b}. SQM may have been produced shortly after
the Big Bang and may have survived as remnants.
SQM may contribute to the cold Dark Matter (DM) in the Universe.
\par The main energy loss for low velocity nuclearites is due to elastic or quasi-elastic
collisions with ambient atoms. The loss  is large;
therefore nuclearites may be easily detected in scintillators and
CR39 NTDs. Nuclearites should have typical galactic
velocities, $\beta$$\sim$$10^{-3}$, and for masses $>$0.1 g could
traverse the Earth.
\par Most nuclearite searches were made as
byproducts of CR MM searches; the flux limits  are similar to those
for MMs.
Direct flux limits for nuclearites come from 
large area experiments with CR39 NTDs; Mt. Norikura at 2770 m
a.s.l.~\cite{Experiments}; at the depth of $\sim$$10^4$~g~cm$^{-2}$ in
the Ohya mine~\cite{Experiments}; MACRO, at a depth of 3700 hg~cm$^{-2}$, used also liquid scintillators \cite{macro-nucl}. 
Experimental limits for heavy nuclearites are at the level of
those presented in Fig. 7b for GUT MMs: $\sim$$1.4\times 10^{-16}$ cm$^{-2}$ s$^{-1}$ sr$^{-1}$. For Intermediate Mass
Nuclearites the limits are at the level indicated in Fig. 7a. 
For small nuclearites, $A<8000$, the
predicted flux in the cosmic radiation is expected to increase with
decreasing mass.
The present status of the search for galactic nuclearites is summarized in
 Fig. 10 \cite{Sahnoun, Cecchini2008b}. Indirect searches may yield
lower limits, but they are affected by systematic uncertainties. Some
exotic cosmic ray events were interpreted as due to incident
nuclearites, f. e. the ``Centauro'' events and the anomalous massive
particles, but the interpretation is not
unique~\cite{polacchi}. 
\par
{\it Q-balls} should be aggregates of squarks $\tilde{q}$, sleptons
$\tilde {l}$ and Higgs fields~\cite{qballs}. The scalar condensate
inside a Q-ball core has a global baryon number Q (and may be also a
lepton number). Protons, neutrons and electrons may be
absorbed in the condensate.
There may be neutral and charged Q-balls: Supersymmetric
Electrically Neutral Solitons (SENS) (more massive and may
catalyse proton decay); SENS may obtain a positive electric charge
when absorbing a proton in their interaction with matter yielding SECS
(Supersymmetric Electrically Charged Solitons), which have a core
electric charge and lower masses; the Coulomb barrier 
may prevent the capture of nuclei. SECS have only integer charges   
because they are color singlets. Fig. 11 \cite{Cecchini2008b} shows sketches
of SECS and SENS. A SENS which enters the Earth atmosphere could
absorb a nitrogen nucleus and become a SECS with charge
$Z_Q$=7. 
Q-balls could be cold DM candidates. SECS with $\beta$$\sim$$ 10 ^{-3}$
and  $M_Q < 10^{13}$ GeV may reach an underground detector from  
above. SENS may be detected by their large 
emission of pions; SECS may be detected by scintillators, NTDs and
ionization detectors.
Fig. 12 shows the present status of the searches for
galactic charged Q-balls with a net charge  $Z_Q=1$.
\vspace{-0.3cm}

\section{Conclusions. Outlook}\label{sec:6}

The NTDs CR39 and Makrofol were calibrated with different ions of different
energies. For each type of detector a unique curve of $p$ vs REL describes their response. 
\par
The total fragmentation cross sections for low energy ions on different targets 
do not show any observable energy dependence and are in agreement 
with similar data in the literature.
\par
Direct and indirect accelerator searches for classical Dirac MMs 
placed limits for $m_M$$\leq$800 GeV. Future improvements may come from LHC experiments \cite{Abbiendi}. 
Searches performed for GUT poles in the penetrating 
cosmic radiation yielded 90\% CL flux limits at the level of $\sim$$1.4 \times 10^{-16} $~cm$^{-2}$~s$^{-1}$~sr$^{-1}$ for 
$\beta \ge 4 \times 10^{-5}$. 
Present limits on IMMs with high $\beta$, in
the downgoing cosmic radiation are at the level of
$1.3\times 10^{-15}$~cm$^{-2}$ s$^{-1}$ sr$^{-1}$. 
 As a byproduct of GUT MM searches some experiments obtained stringent
limits on nuclearites, strangelets and Q-balls. 

In the past, a number of monopole and of other exotic candidates 
were thought to have been observed
and some results were published \cite{polacchi, exotic}. But they were not
confirmed and most of them are now neglected. 
In 2006 the SLIM experiment faced a problem of this type when analysing the
top faces of the top CR39 layers. A sequence
of etch-pits was found along a $\sim$20 cm
line; each one of them looked complex and very
different from usual ion tracks. Since the ``candidate event'' was rather
peculiar, a thorough study was made in all the sheets of
the module, and in all NTD sheets in the modules 
within a $\sim$1 m distance. Short soft etching periods were used so as to follow 
the evolution of the etch-pits. It was concluded that they originated
in a rare manufacturing defect involving 1 m$^2$ of CR39 \cite{Balestra2008b}.
\par
We measured the radon concentration in the houses of the city of Bologna, in the Gran Sasso Underground Laboratory and
 in some thermal sites. In the first 2 cases the radon level was globally low, and changed with the floor and ventilation 
\cite {radon1}.
At the 2008 
24$^{th}$ Int. Conf. on Nuclear Tracks in Solids, in Bologna, new results were presented in Radiation 
Environment Monitoring (mainly radon), Neutron Dosimetry and Medical Applications (see proceedings in 
Radiation Measurements). \\

We acknowledge several discussions with many colleagues. We thank Drs. M. Errico, M. Giorgini and I. Traor\'e for their 
cooperation.


\vspace{-0.3cm}

\begin{thebibliography}{99}
\bibitem{Durrani}  S.A. Durrani et al., Solid State Nuclear Track Detection, Pergamon Press, Oxford (1987).
\bibitem{Nikezic} D. Nikezic et al., Material Science Eng. R46 (2004) 51.
\bibitem{Uchihori} Y. Uchihori et al., J. Radiat. Res. 43 (Suppl. S81-5) (2002).
\bibitem{Kodaira} S. Kodaira et al., Jpn. J. Appl. Phys. 43 (2004) 6358.
\bibitem{Ambrosio2002a} M. Ambrosio et al., Eur. Phys. J. C25 (2002) 511; Nucl. Instrum. Meth. A486 (2002) 663.
\bibitem{Balestra2008} S. Balestra et al., Eur. Phys. J. C 55 (2008) 57;  hep-ex/0506075;  hep-ex/0602036. \newline 
V. Togo and I. Traor\'e, arXiv:0811.2885 [physics.ins-det].
\bibitem{Cecchini2005a} S. Cecchini et al., astro-ph/0510717 (2005). 
\bibitem{Cecchini2005b} S. Cecchini et al., Radiat. Meas. 40 (2005) 405.
\bibitem{Derkaoui} J. Derkaoui et al., Astropart. Phys. 10 (1999) 339. D. Bakari et al., hep-ex/0004019.
\bibitem{Giacomelli2007} G. Giacomelli et al., hep-ex/0702050 (2007).
\bibitem{Manzoor2007} S. Manzoor et al., Nucl. Phys. B (Proc. Suppl.) 172 (2007) 296.
\bibitem{Medinaceli} E. Medinaceli, arXiv:0811.1111 [hep-ex].  
\bibitem{Sahnoun} Z. Sahnoun, 24th Int. Conf. Nucl. Tracks in Solids, (2008) Bologna, Radiat. Meas.
\bibitem{Cecchini2008} S. Cecchini et al., Nucl. Phys. A 807 (2008) 206.
\bibitem{Giorgini} M. Giorgini, arXiv:0812.0236 [nucl-ex]; arXiv:0812.0685 [physics.ins-det]. 
\bibitem{Togo} V. Togo et al., Nucl. Instrum. Meth. A 580 (2007) 58.
\bibitem{Patrizii} L. Patrizii et al., Nucl. Tracks Radiat. Meas. 19 (1991) 641.
\bibitem{Vilela} E. Vilela et al., Radiat. Meas. 31 (1999) 437.
\bibitem{Balestra2007} S. Balestra et al., Nucl. Instrum. Meth. B 254 (2007) 254.
\bibitem{Noll} A. Noll et al., Nucl. Tracks Radiat. Meas. 15 (1988) 265.
\bibitem{Giacomelli98} G. Giacomelli et al., Nucl. Instrum. Meth. A 411 (1998) 41.
\bibitem{Cecchini93} S. Cecchini et al., Astropart. Phys. 1 (1993) 369.
\bibitem{Bradt} H.L. Bradt et al., Phys. Rev. 77 (1950) 54. S. Cecchini et al., Nucl. Phys. A707 (2002) 513.
\bibitem{gg+lp} G. Giacomelli et al., hep-ex/011209; hep-ex/0302011;   
  hep-ex/0211035; hep-ex/0506014.
\bibitem{ruzicka}J. Ruzicka and V.P. Zrelov JINR-1-2-80-850 (1980).
\bibitem{biblio} G. Giacomelli et al., hep-ex/0005041.
\bibitem{mm_macro} M. Ambrosio et al., Phys. Lett. B406 (1997) 249;
  Astropart. Phys. 18 (2002) 27.
\bibitem{Experiments} S. Orito et al.,
  Phys. Rev. Lett. 66 (1991) 1951. Yu.F. Novoseltsev et al., Nucl. Phys. B151 (2006) 337.  
V. Aynutdinov et al.,astro-ph/0507713. A. Pohl et al., astro-ph/0701333.
\bibitem{price} P. B. Price, Phys. Rev. D38 (1988) 3813.
 D. Ghosh et al., Europhys. Lett. 12 (1990) 25.
\bibitem{lazaride} T. W. Kephart and Q. Shafi, Phys. Lett. B520 (2001) 313. 
\bibitem{nucleariti} E. Witten, Phys. Rev. D30 (1984) 272.
  A. De Rujula and S. Glashow, Nature 31 (1984) 272.
\bibitem{Cecchini2008b}  S. Cecchini et al., arXiv:0805.1797 [hep-ex].
\bibitem{macro-nucl} M. Ambrosio et al.,
  Eur. Phys. J. C13 (2000) 453. S. Ahlen et al., Phys. Rev. Lett. 69 (1992) 1860.
\bibitem{polacchi} M. Rybczynski et al., hep-ph/0410064. D. P. Anderson et al., astro-ph/0205089
\bibitem{qballs} S. Coleman, Nucl. Phys. B262 (1985) 293. 
  A. Kusenko et al., Phys. Lett. B418 (1998) 46.
\bibitem{Abbiendi} G. Abbiendi et al., Phys. Lett. B663 (2008) 37.
\bibitem{exotic} P.B. Price et al., Phys. Rev. Lett 35 (1975) 487;
  Phys. Rev. D18 (1978) 1382.
\bibitem{Balestra2008b} S. Balestra et al., arXiv:0802.2056 [hep-ex]. 
\bibitem{radon1} G. Giacomelli et al., Il Nuovo Saggiatore 4 (1990) 5; Acqua Aria 4 (1990) 371.
 C. Arpesella et al., Health Phys. 72 (1997) 629.
\end{thebibliography}
\end{document}